\documentclass[3p]{elsarticle}

\usepackage{amssymb}
\usepackage[normalem]{ulem}
\usepackage{amsmath}
\usepackage{float}
\usepackage{todonotes}
\usepackage[colorlinks=true, allcolors=blue]{hyperref}
\usepackage{enumitem} 
\usepackage{soul}
\usepackage{empheq}
\usepackage{sidecap}
\usepackage{xcolor}
\usepackage{multirow}
\usepackage{graphbox}
\usepackage{booktabs}
\usepackage{longtable}
\usepackage{verbatimbox}
\usepackage{array}
\usepackage{fancyhdr}
\usepackage{booktabs}
\usepackage{tabularx}
\usepackage{subcaption}
\usepackage{bm}
\usepackage{graphicx}
\usepackage{subcaption}
\usepackage{xcolor}
\usepackage[percent]{overpic}
\usepackage{lineno}

\journal{}

\makeatletter
\def\ps@pprintTitle{%
  \let\@oddhead\@empty
  \let\@evenhead\@empty
  \let\@oddfoot\@empty
  \let\@evenfoot\@empty
}
\makeatother

\begin{document}

\begin{frontmatter}



\title{The effect of interactions on elastic cavitation}

\author[UNH]{Ali Saeedi}
\author[Purdue]{S. Chockalingam}
\author[UNH]{Mrityunjay Kothari\corref{cor1}}
\cortext[cor1]{Corresponding author: mrityunjay.kothari@unh.edu}
\address[UNH]{Department of Mechanical Engineering, University of New Hampshire, Durham, NH 03820, USA}
\address[Purdue]{Department of Mechanical Engineering, Purdue University, West Lafayette, IN 47907, USA}
\begin{abstract} 
Cavitation refers to the sudden, unstable expansion of a defect or cavity within a material in response to applied loads, when the loads reach a critical threshold. 
It is widely recognized as a common failure nucleation mechanism in soft and biological materials. 
For an isolated cavity in the bulk of an incompressible neo-Hookean solid loaded by remote hydrostatic tension, the classical cavitation pressure is well established as $p_{\rm c, bulk}=2.5\mu$, where $\mu$ is the shear modulus. 
However, in realistic settings the cavitation threshold is influenced by additional factors, including proximity to interfaces and interactions with neighboring cavities. 
Interface interaction effects are particularly relevant in modern multi-material systems and additively manufactured structures, where defects frequently occur near material boundaries. 
Meanwhile, cavity-cavity interactions become important in materials exhibiting finite porosity, such as foams, porous solids, and phase-separating polymers.

In this study, we characterize the effect of cavity interaction on cavitation pressure for (i) a nearby rigid interface and (ii) a neighboring identical cavity. 
For cavities near a rigid interface, our analysis shows that the cavitation pressure increases as the initial cavity-interface distance decreases, starting from the bulk value $p_{\rm c, bulk}$ for a distant cavity and approaching the cavitation pressure value for a defect situated at an interface ($p_{\rm c, int}\approx3.5\mu$) as the cavity approaches the interface boundary. 
In contrast, interacting cavities exhibit a non-monotonic dependence of the cavitation pressure on the initial inter-cavity distance $d$: the threshold approaches the bulk value $p_{\rm c, bulk}$ for distant cavities and reaches a maximum of {$p_{\rm c}\approx 2.67\mu$} at {$d\approx 3.8R$}, where $R$ is the initial cavity radius. 
These results provide, for the first time, a quantitative estimate of cavitation pressure for these two interaction-effect extensions of the classical cavitation problem and will help guide modeling of cavitation-driven processes in porous, biological, and multi-material solids.
\end{abstract}


\end{frontmatter}

\section{Introduction}
\label{sec:introduction}

Cavitation is a mechanical instability in which small pre-existing voids 
or defects embedded in a solid expand suddenly and unstably once an applied load crosses a critical threshold. It is a critical failure nucleation mechanism in soft and biological materials \cite{barney2020cavitation} as well as metals \cite{huang1991cavitation}. The phenomenon 
was first systematically characterized in rubber-like materials by Gent and 
Lindley \cite{GentLindley1959}, who showed that spherical cavities embedded 
in an incompressible neo-Hookean solid grow without bound when the remotely 
applied hydrostatic tension reaches the critical value $p_{\rm c,bulk} = 
2.5\mu$, where $\mu$ is the shear modulus. This result was later 
reinterpreted by Ball \cite{Ball1982} as a bifurcation of the equilibrium 
equations of nonlinear elasticity: a solid ball that is perfectly defect-free 
may spontaneously open a cavity at its center when the applied load is 
sufficiently large, yielding the same critical pressure as the defect-growth 
picture.\\

Since these foundational studies, cavitation has been investigated 
extensively across a wide range of material systems and loading scenarios. 
The role of material compressibility was examined by Horgan and Abeyaratne 
\cite{horgan1986bifurcation} for foam rubberlike (Blatz--Ko) materials, with 
further explicit radially symmetric solutions for compressible nonlinear 
elastic solids provided by Horgan \cite{horgan1992void}. Stuart 
\cite{STUART198533}, Sivaloganathan \cite{sivaloganathan1986uniqueness}, and 
Meynard \cite{meynard1992existence} addressed existence and uniqueness of cavitating 
equilibria for broader classes of compressible materials. 
Further extensions of radially symmetric cavitation to anisotropic materials 
were studied by Antman and Negr\'on-Marrero \cite{antman1987remarkable} and by 
Polignone and Horgan \cite{polignone1993cavitation}, while dynamic \cite{pericak1988nonuniqueness} and viscoelastic \cite{cohen2015dynamic,kumar2017some} effects have also been investigated. A 
review of radially symmetric cavitation in nonlinear elastic solids can 
be found in \cite{horgan1995cavitation}.\\

Hou and 
Abeyaratne \cite{hang1992cavitation} extended cavitation analysis to non-symmetric 
loading by constructing an approximate cavitation surface in full principal-stress 
space, showing that the hydrostatic pressure criterion provides a useful but 
imperfect approximation when the loading deviates from purely volumetric tension. 
Lopez-Pamies 
\cite{LopezPamies2009} later derived a closed-form cavitation criterion for 
compressible, isotropic, hyperelastic solids under general plane-strain 
loading using a linear-comparison variational method.
Lopez-Pamies and co-workers \cite{lopez2011cavitationI,lopez2011cavitationII} subsequently recast cavitation as a 
homogenization problem, recovering Ball's classical result as a special case 
and extending it to arbitrary three-dimensional loading conditions, 
compressible and anisotropic materials, and general random distributions of 
pre-existing defects. In addition to vacuous cavities and defects, non-symmetric cavitation triggered by inclusions growing inside the material was studied by Bonavia and co-workers \cite{bonavia2026nonlinear}, where the resistance pressure faced by the growing inclusion saturates at a critical value with increasing volume of inclusion.\\ 

While these works provide increasingly general descriptions of cavitation, most remain focused on isolated  growth of cavities in a perfect elastic medium. However, cavities rarely grow in isolation in
biological and modern engineering materials. Soft composites contain inclusions and 
reinforcement particles, where growth of cavities  near material interfaces affects cavitation and failure
\cite{GentPark1984,cho1988cavitation}. Similarly in porous solids and biological tissues, and phase separating polymers, growing cavities can interact with each other which can affect cavitation and failure. In porous 
elastomers for example, Lopez-Pamies and Ponte Castañeda \cite{LopezPamies2007} showed 
that interactions among cavities encoded in the homogenized constitutive response 
can trigger macroscopic instabilities. Thus it is critical to study how cavitation is affected by the interaction of growing cavities with nearby interfaces and other growing cavities to robustly model and understand the behavior of porous, biological, and multi-material solids.\\

There are a limited number of cavitation studies that focus on interaction effects. Kang and coworkers \cite{Kang2018}  studied cavity expansion in inhomogeneous soft solids and demonstrated  that proximity to a planar interface between two  materials of differing moduli can significantly alter the cavitation response.  In an earlier study, Henzel and co-workers \cite{Henzel2022} analyzed cavitation arising from a defect that exists at the interface between an elastic body and a bonded rigid substrate and established
that the critical pressure rises to $p_{\rm c,int} \approx 3.5\mu$---a 
forty-percent increase over the bulk value attributable entirely to the 
constraint imposed by the rigid boundary. The framework of Lopez-Pamies and co-workers \cite{lopez2011cavitationI} can be used to account for inter-cavity interactions but only in an average homogenized sense through an assumed distribution of voids.  There are no studies so far that have analyzed how the proximity of growing cavities to nearby interfaces and other growing cavities affects the cavitation limit. The current work addresses this gap. \\

In this manuscript, we study the effect of interactions on cavitation while neglecting surface tension and fracture physics (to isolate the effect of interactions). 
Specifically, we focus on two fundamental cases---the interaction of a growing cavity with a nearby rigid interface, and, separately, with a nearby identically growing cavity. 
Thus, we consider an incompressible neo-Hookean solid under remote 
hydrostatic tension loading and analyze two configurations: (i) a single spherical 
cavity with radius $R$ at a distance $d$ from a nearby planar rigid interface, and (ii) two identical spherical cavities of radii $R$
separated by distance $2d$. Our prior work \cite{Saeedi2D, Saeedi2025} characterized the energy landscape and configurational forces in the latter setting, but here we focus on the cavitation limit.\\

The paper is organized as follows. 
We begin by presenting the results of the study in \S\ref{sec:results}, followed by a detailed 
description of the problem and solution setup used to obtain the results in \S\ref{sec:problem_setup}. 
The computational framework is outlined in \S\ref{sec:comp_framework}, and we end with concluding remarks in \S\ref{sec:conclusion}.

\begin{figure}[h]
    \centering
    \includegraphics[width=0.6\linewidth]{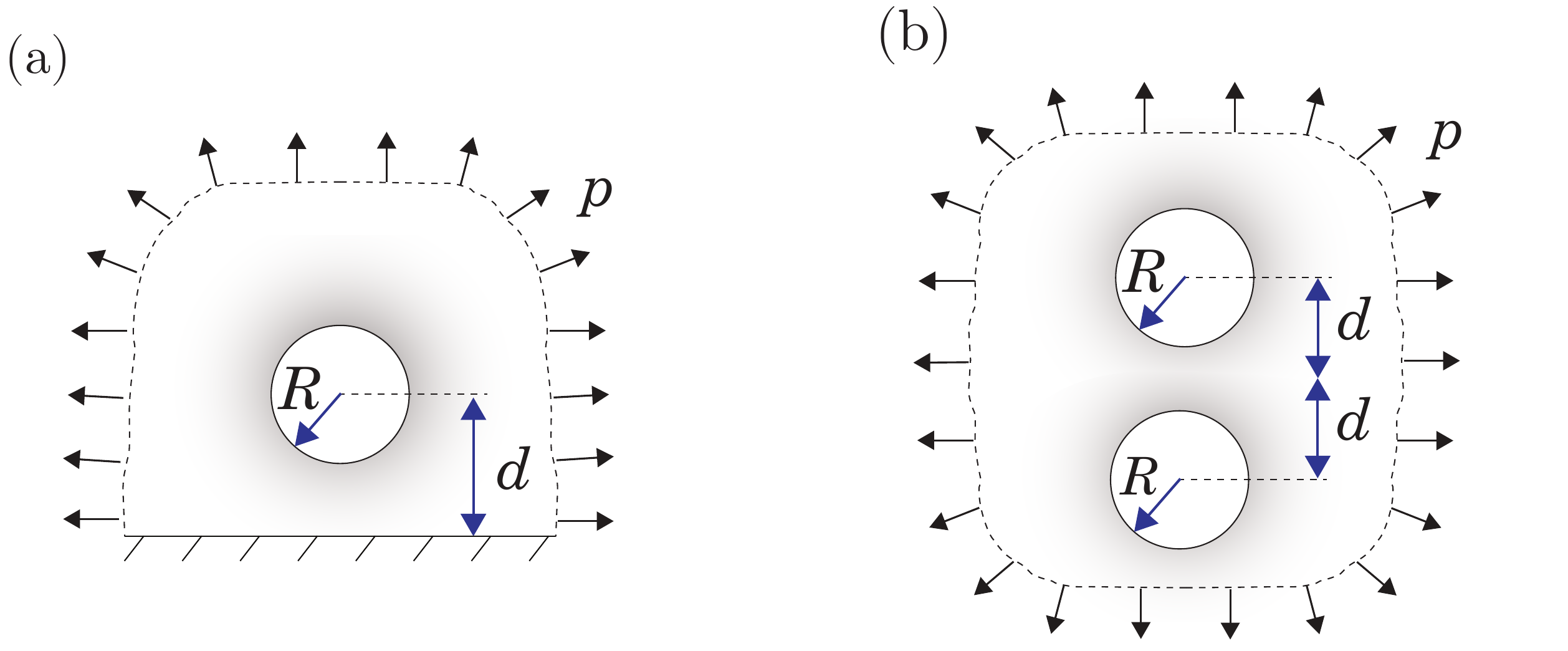}
    \vspace{-10pt}
    \caption{{Axisymmetric schematic of the cavitation problems being studied: (a) cavity-interface interaction, and (b) cavity-cavity interaction. The cavities are under remote hydrostatic tension loading.}}
    \label{fig:Fig1}
\end{figure}

\section{Results } \label{sec:results}

\begin{figure}[h]
    \begin{subfigure}[t]{0.515\linewidth}
    \begin{overpic}[width=\linewidth]{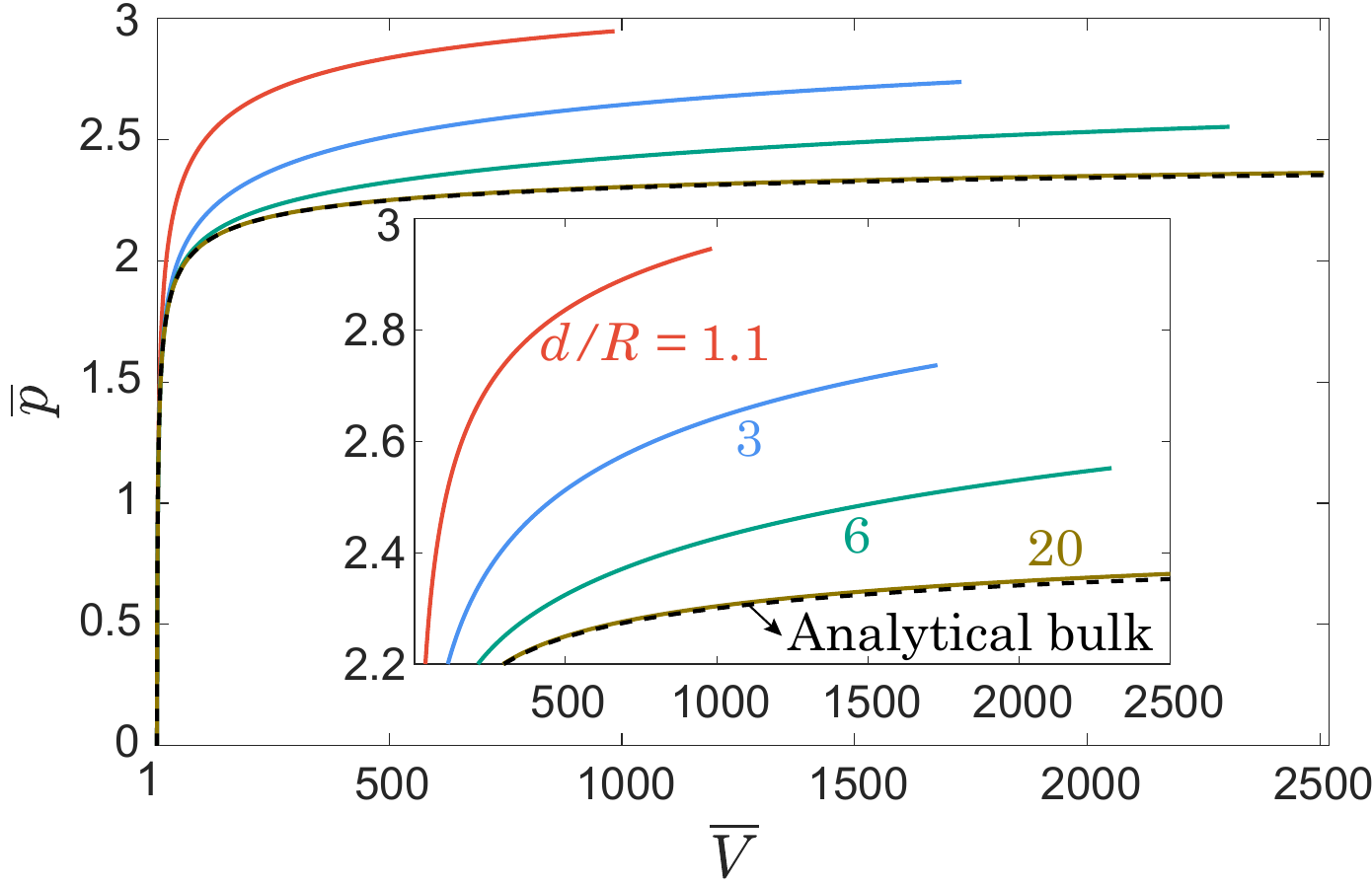}
    \put(0,63){\text{(a)}}
    \end{overpic}
    \phantomcaption
    \label{fig:Fig2a}
    \end{subfigure}
    \hfill
    \begin{subfigure}[t]{0.45\linewidth}
    \begin{overpic}[width=\linewidth]{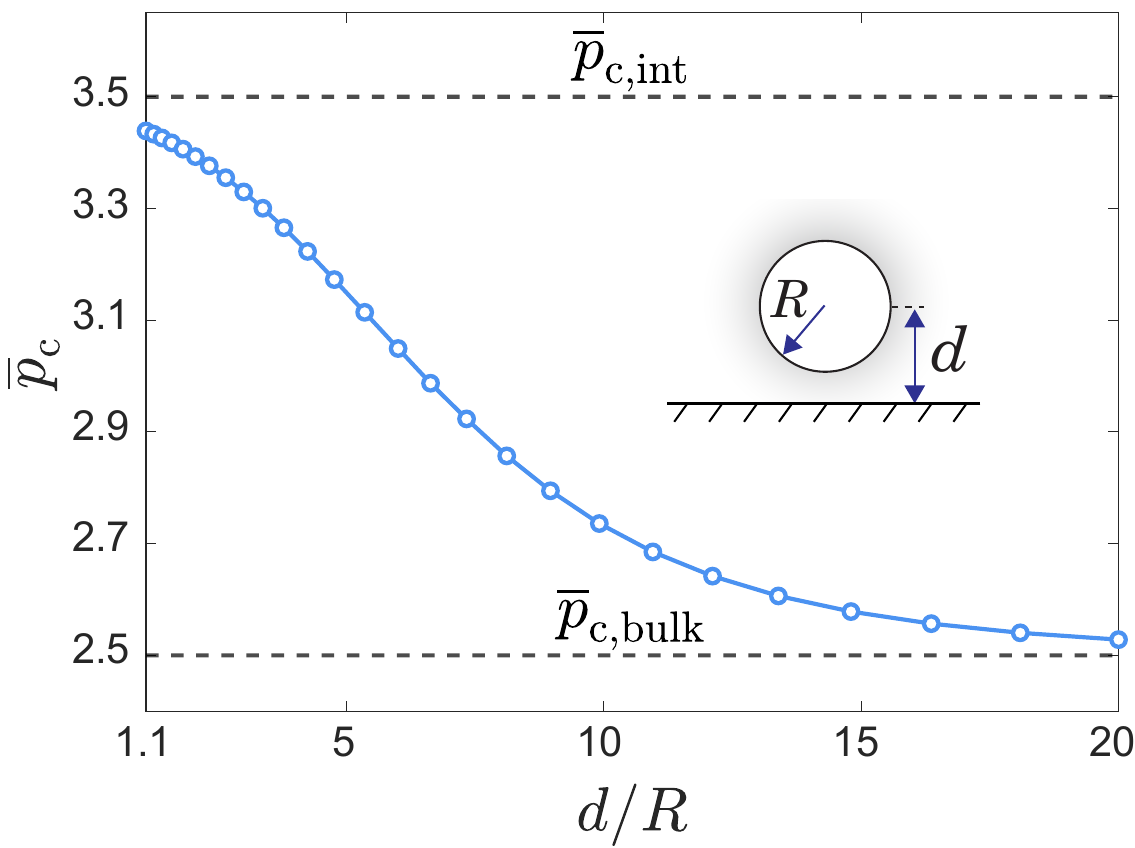}
    \put(0,72){\text{(b)}}
    \end{overpic}
    \phantomcaption
    \label{fig:Fig2b}
    \end{subfigure}
    \vspace{-20pt}
    \caption{Cavitation near a rigid interface: (a) dimensionless pressure--volume curves for different dimensionless cavity--interface distances $d/R=\{1.1, 3, 6, 20\}$. The dashed black curve shows analytical bulk cavitation case (eq. \eqref{eq:Pc_analytical}); (b) dimensionless cavitation pressure as a function of $d/R$.} 
    \label{fig:fixedBC_combined}
\end{figure}

Proximity to an interface or another cavity (Fig. \ref{fig:Fig1}) affects the cavitation pressure relative to its bulk value.
To quantitatively describe these effects, we introduce the following notation. 
The dimensionless pressure is defined as $\overline{p}=p/\mu$, where $p$ is the remotely applied pressure and $\mu$ is the shear modulus of the material.
The dimensionless volume ratio is defined as $\overline{V}=V/V_0$, where $V_0$ is the initial, undeformed cavity volume, and $V$ is its deformed volume. 
The critical value of pressure $p$ which causes unstable cavity expansion is the cavitation pressure $p_{\rm c}$, and $\overline{p}_{\rm c} = p_{\rm c}/\mu$ is the dimensionless cavitation pressure.
As a reference value, the bulk cavitation pressure for a spherical cavity in an infinite, incompressible neo-Hookean medium is \(\overline{p}_{\rm c, \rm bulk}=2.5 \). The cavitation pressure for expansion of a defect located exactly at the interface between the elastic body and a bonded rigid substrate was established in a prior study \cite{Henzel2022} as 
$\overline{p}_{\rm c, int}\approx3.5$, also called the interfacial cavitation pressure.
The key results of this study are as follows:

\paragraph{Cavity near a rigid interface in a semi-infinite medium} Fig.~\ref{fig:Fig2a} shows the pressure-volume relation for a sequence of distances (expressed in the dimensionless form $d/R$ where $R$ is the radius of the cavity and $d$ is the distance of its center from the rigid interface in the undeformed configuration) and Fig.~\ref{fig:Fig2b} shows the corresponding post-processed cavitation pressures.
The cavitation pressure increases monotonically as the undeformed cavity location gets closer to the interface, attaining {$\overline{p}_{\rm c} \approx 3.44$ at $d/R = 1.1$}---a significant increase from the bulk cavitation pressure (which is the asymptotic limit of the cavitation pressure for $d/R \to \infty$). As the distance between the cavity and the interface increases, the pressure-volume curve approaches that of the bulk cavitation case (dashed black curve in Fig.~\ref{fig:Fig2a}, defined in eq. \eqref{eq:Pc_analytical}).\\

\begin{figure}[h]
    \centering
    \begin{subfigure}[t]{0.515\linewidth}
    \begin{overpic}[width=\linewidth]{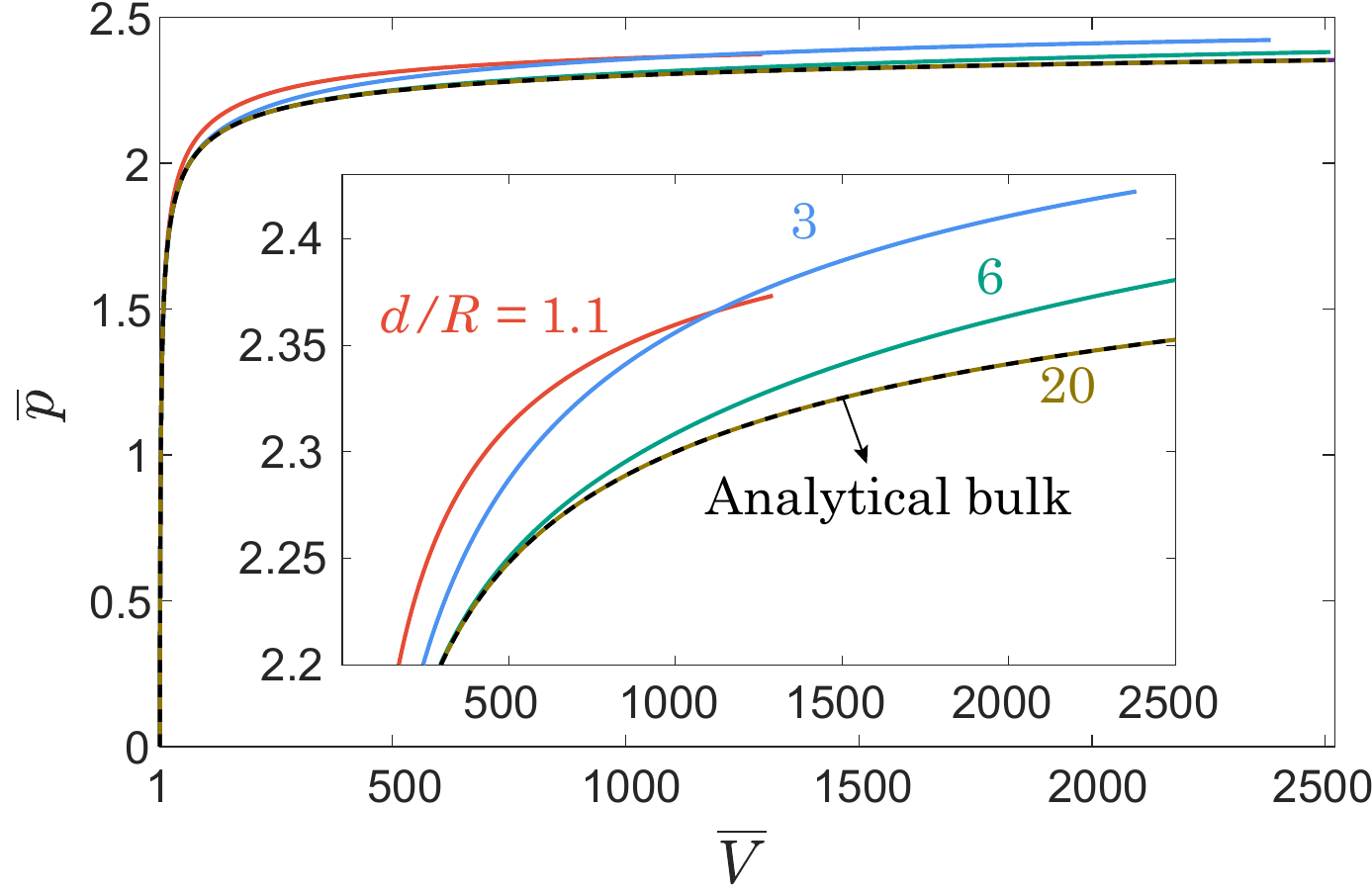}
    \put(0,63){\text{(a)}}
    \end{overpic}
    \phantomcaption
    \label{fig:Fig3a}
    \end{subfigure}
    \hfill
    \begin{subfigure}[t]{0.45\linewidth}
    \begin{overpic}[width=\linewidth]{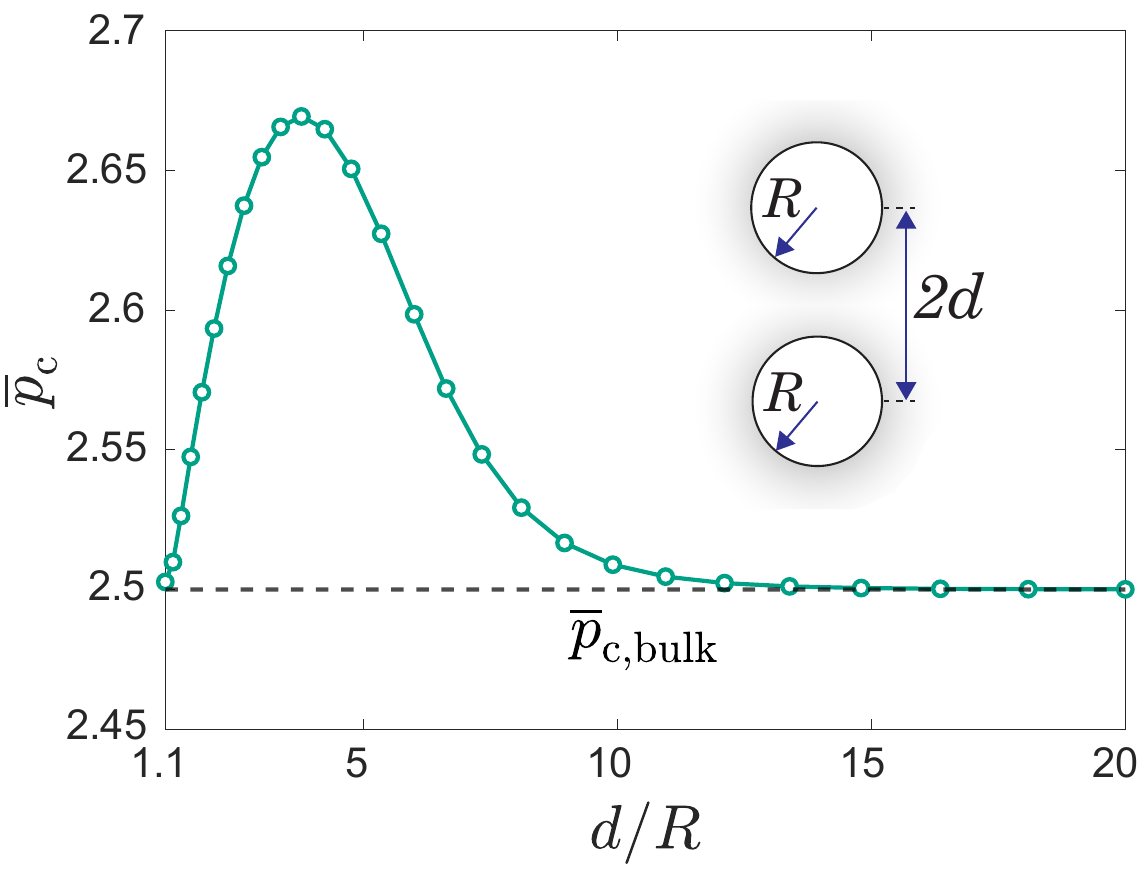}
    \put(0,72){\text{(b)}}
    \end{overpic}
    \phantomcaption
    \label{fig:Fig3b}
    \end{subfigure}
    \vspace{-20pt}
    \caption{Cavitation of two interacting cavities: (a) dimensionless pressure-volume curves of two interacting cavities for different dimensionless inter-cavity half-distances $d/R=\{1.1, 3, 6, 20\}$. The dashed black curve shows analytical bulk cavitation case (eq. \eqref{eq:Pc_analytical}); (b) dimensionless cavitation pressure as a function of $d/R$. }
    \label{fig:symmBC_combined}
\end{figure}

\paragraph{Two cavities in an infinite medium}
The cavitation pressure varies non-monotonically with the undeformed inter-cavity half-distance $d$, as shown in Fig.~\ref{fig:Fig3b}.
A peak is observed at {$d/R\approx 3.8$} where {$\overline{p}_{\rm c} \approx 2.67$}, highlighting a maximal cavity interaction effect; for large $d/R$, the response approaches the single-cavity limit $\overline{p}_{\rm c}=2.5$ (dashed black curve in Fig.~\ref{fig:Fig3a}).\\ 

\begin{figure}[h]
    \centering
    \includegraphics[width=0.95\linewidth]{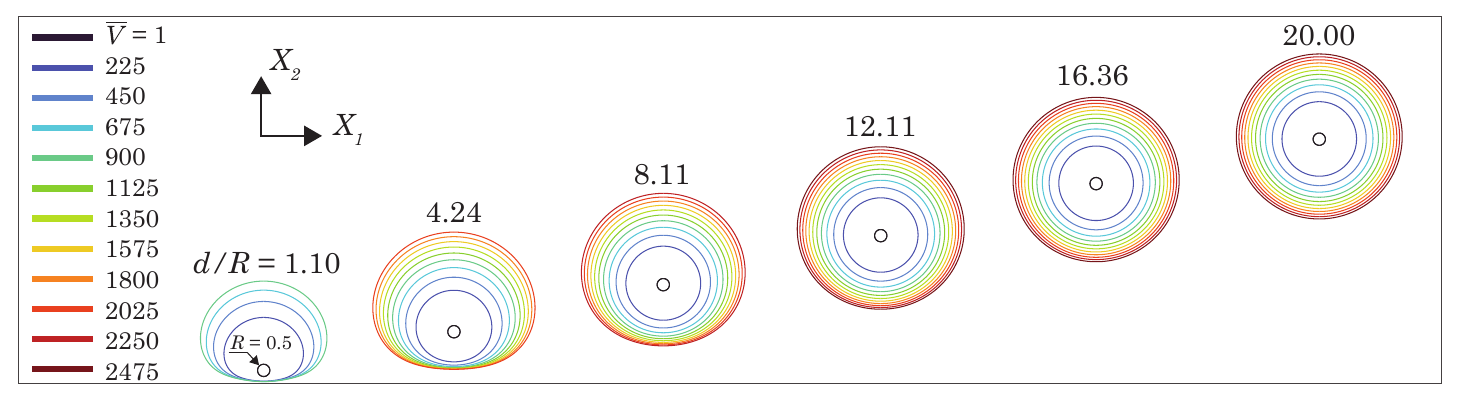}

    \includegraphics[width=0.95\linewidth]{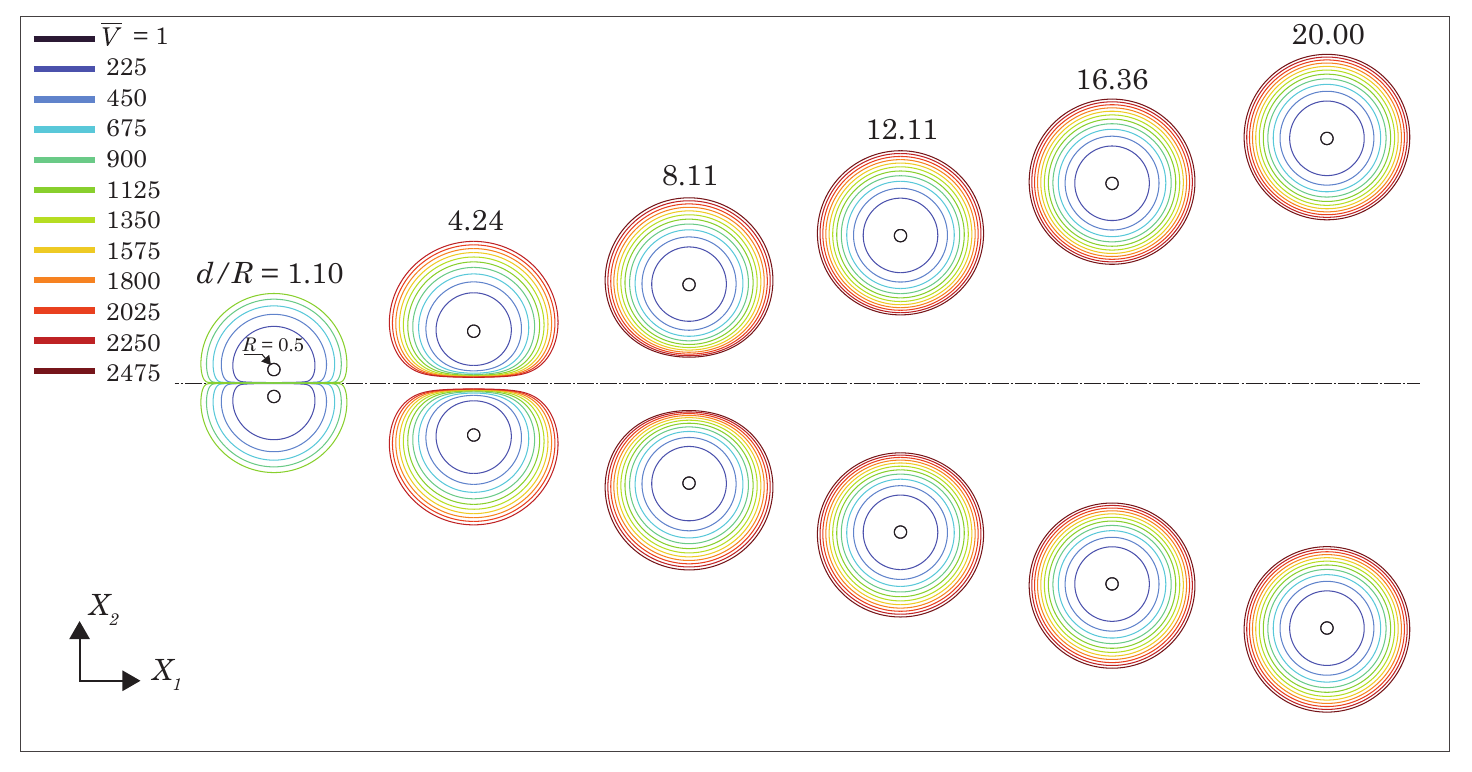}
    \caption{Deformed cavity shapes in (top) cavity-interface interaction and (bottom) cavity-cavity interaction problems for different dimensionless volumes.
    }
    \label{fig:deformed_combined}
\end{figure}

The deformed shapes of the cavity have been plotted at several discrete $\overline{V}$ values for both the cavity-interface and cavity-cavity interaction problems in Figure \ref{fig:deformed_combined}.
It is worth pointing out that the enhancement effect in cavitation pressures found in this study is meaningful for physical systems.
While one may contend that, in theory, other ``isolated'' defects  might cavitate earlier when the pressure reaches the (lower) bulk cavitation value, in real physical systems, surface energy effects impose severe penalties on the nucleation and growth of small defects \cite{dollhofer2004surface,Li2025}.
We assume that the cavities considered in our study are the most relevant size-critical defects in the system.

\section{Problem Setup }
\label{sec:problem_setup}

\subsection{Geometry and Kinematics}
\begin{figure}[h]
    \centering
    \includegraphics[width=1\linewidth]{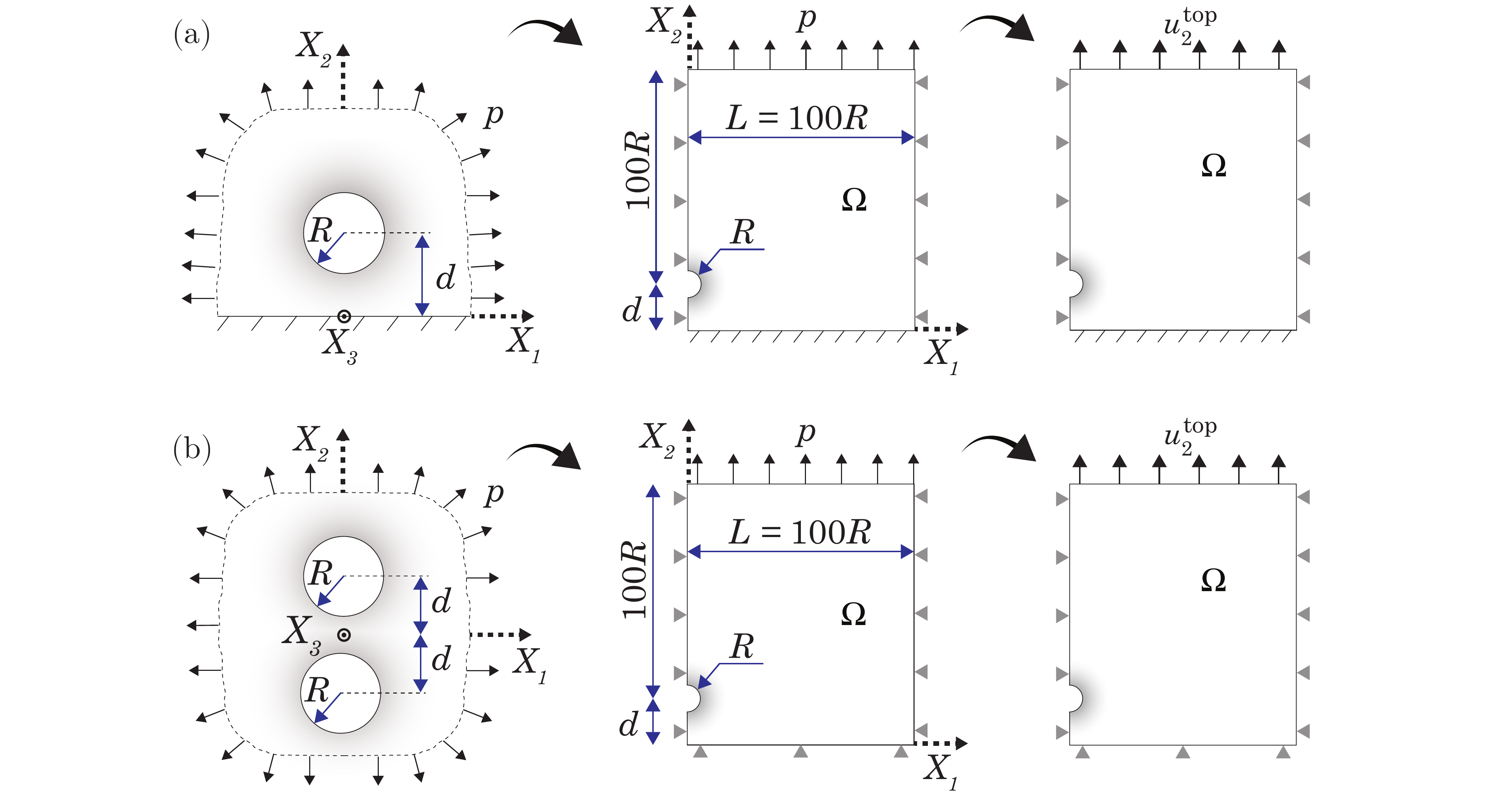}
    \vspace{-20pt}
    \caption{Axisymmetric  schematic of the cavitation problem and its simulation model for the case of  (a) cavity-interface interaction, and (b) cavity-cavity interaction. Both load-controlled and displacement-controlled representations are shown.
    }
    \label{fig:geometry_combined}
\end{figure}

\noindent We investigate two problems in this work: i) a spherical cavity near a rigid interface in a semi-infinite medium, and ii) two equi-sized spherical cavities in an infinite medium.
In both problems, the system is subjected to remote pressure (hydrostatic tension) loading\footnote{The findings from this problem setup are also applicable to physical situations where the cavities are instead internally pressurized (such as from biological \cite{li2022nonlinear,chockalingam2024large,senthilnathan2024understanding, kothari2023crucial, zhang2021morphogenesis, nijjer2023biofilms} and other growth \cite{kothari2020effect,bonavia2026nonlinear} processes, thermo-chemical processes in energetic materials \cite{chockalingam2022thermo}, or through injected fluid in cavity-based viscoelastic characterization techniques \cite{chockalingam2021probing,li2025cylindrical}) since we can show equivalence in the displacement fields of the remote loading and internal loading problems \cite{henzel2022reciprocal}.}.
These are both axisymmetric problems, whose 2D representations are shown in Figure~\ref{fig:geometry_combined}, with $X_2$ being the axis of symmetry. In numerical simulations, the remote boundaries are implemented using large finite dimensions; the additional symmetry about the $X_1$ axis in the interacting-cavities problem allows us to reduce its geometry to the same as that of the cavity-interface interaction problem except that the bottom boundary is a symmetry plane instead of a rigid interface.

\paragraph{Approach} The cavitation problem is a load-controlled problem.
However, since we do not know \textit{a priori} the cavitation pressure to approach the limit in a controlled manner, the load-controlled problem is not very tractable.
We employ a displacement-controlled approach to alleviate this issue.
Given the infinite size of the domain, the displacements at the remote boundaries are expected to be uniform.
The volume change of the cavity can then be directly related to the imposed uniform displacements since the material is incompressible.
The equivalent displacement-control problem is shown in Figure~\ref{fig:geometry_combined}. \\

The volume change at the top boundary must equal the volume change for cavity, which determines the displacement at the top boundary as
\begin{equation}
    u_2^{\mathrm{top}} = \frac{V-V_0}{\pi L^2},
    \label{eq:u2_top}
\end{equation}
where $L$ is the length of the top edge of the simulation domain (see Fig.~\ref{fig:geometry_combined}).

\subsection{Material Model}
\noindent The medium $\Omega$ is modeled as incompressible neo-Hookean. 
Strain energy density for an incompressible neo-Hookean material is 
\begin{equation}
    W = \frac{\mu}{2} (I_1 - 3),
\end{equation}
where $\mu$ is the shear modulus of the material, $I_1=\text{tr}(\textbf{B})$ is the first invariant of the left Cauchy-Green deformation tensor $\textbf{B}=\textbf{F}\textbf{F}^T$, and $\textbf{F}$ is the deformation gradient tensor with $\det \textbf{F} = 1.$\\

The analytical dimensionless pressure-volume relation for the growth of a spherical cavity in an infinite incompressible neo-Hookean medium is given as \cite{GentLindley1959,Saeedi2025}
\begin{equation}
    \overline{p} = \frac{1}{2} (1-\overline{V}^{-4/3}) + 2 (1-\overline{V}^{-1/3}).
    \label{eq:Pc_analytical}
\end{equation}
where it is seen that the bulk cavitation limit can be found by setting $\overline{V} \to \infty$ as \(\overline{p}_{\rm c, bulk}=2.5\).

\section{Computational Framework } \label{sec:comp_framework}
\subsection{Discretization and Simulation Setup}
\begin{figure}[h]
    \centering
    \includegraphics[width=0.8\linewidth]{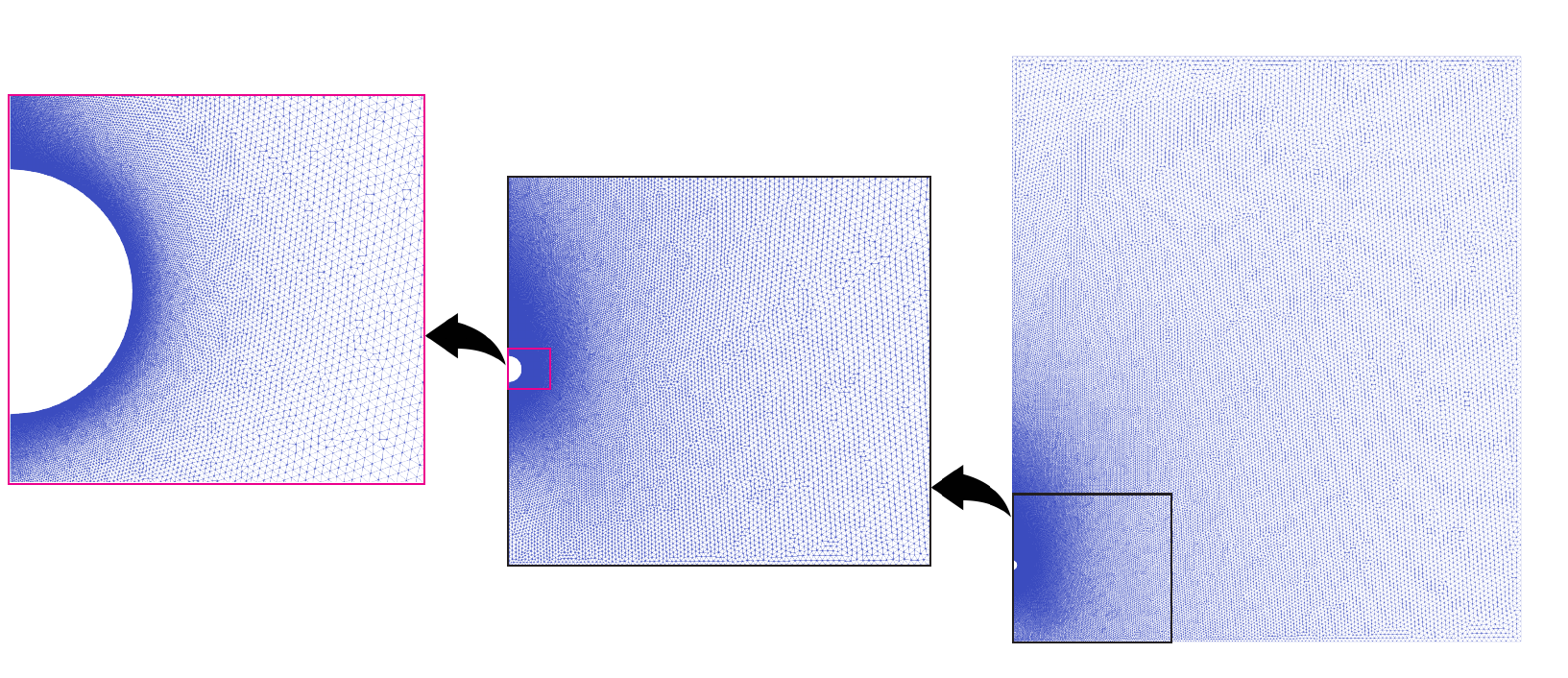}
    \vspace{-10pt}    
    \caption{Spatially adaptive mesh discretization ($d/R=7.5$).}
    \label{fig:discretized}
\end{figure}

\noindent We carry out the simulations using the open source finite element software \texttt{FEniCS} \cite{alnaes2015fenics, logg2012automated, logg2012dolfin}. Incompressibility is enforced via a mixed displacement-pressure ($u-\tilde{p}$) formulation 
using the lowest-order Taylor-Hood element pair (quadratic elements for 
displacement and linear elements for pressure), with a perturbed Lagrangian 
treatment of the volumetric constraint ($J - 1 = \tilde{p}/K$, where $\tilde{p}$ is the Lagrange multiplier 
interpreted as a hydrostatic pressure and $K = 10^{12} \times G$ is a penalty 
parameter with the interpretation of a bulk modulus). The weakform setup in FEniCS parallels the implementation in \cite{senthilnathan2024understanding} (see also \cite{Anand2026}).
 The domain dimensions used in the simulation are shown in Figure \ref{fig:geometry_combined}. 
The mesh is especially refined near the cavity to ensure sufficient accuracy in capturing the large deformations in the vicinity of the cavity. 
Figure~\ref{fig:discretized} shows an example of the meshes employed; here, the minimum element size is $R/250$, and 168758 triangular elements are used to discretize the domain using the software \texttt{Gmsh} \cite{geuzaine2009gmsh}.
Mesh convergence study is presented in \ref{app:convergence}.
The boundary value problems illustrated in Figure \ref{fig:geometry_combined} are numerically solved for both rigid and symmetric interfaces for dimensionless distance values $d/R \in \{$1.10, 1.25, 1.41, 1.60, 1.82, 2.06, 2.33, 2.65, 3.00, 3.37, 3.78, 4.24, 4.76, 5.35, 6.00, 6.63, 7.33, 8.11, 8.96, 9.91, 10.95, 12.11, 13.39, 14.80, 16.36, 18.09, 20.00$\}$. 
The pressure–volume data are then extracted (see \ref{app:convergence}), resulting in the plots shown in Fig.~\ref{fig:Fig2a} and Fig.~\ref{fig:Fig3a}.

\subsection{Cavitation Pressure through Curve Extrapolation}

\noindent Given that the simulations show a monotonic increase in cavity volume with applied pressure, where the pressure appears to asymptote in all cases, the cavitation pressure can be determined by the limiting value of the pressure as $\overline{V} \to \infty$.
While the simulations are able to achieve large values of $\overline{V}$, the pressure-volume curves need to be extrapolated to determine the asymptotic cavitation pressure value. We achieve this by employing a curve-fitting procedure for the pressure-volume curve and then using the fitted curve to estimate the cavitation pressure.

Motivated by the expression for the pressure-volume relation for the bulk cavitation case in eq.~\eqref{eq:Pc_analytical}, we employ polynomial basis functions in terms of the variable $\overline{V}^{-1/3}$ for the fitting. 
Specifically, we choose Laguerre polynomials of degree $m$, given by
\begin{equation}
    L_m(x) = \sum_{k=0}^{m} \binom{m}{k} \frac{(-1)^k}{k!} x^k,
\end{equation}
since they are orthonormal in $x\in [0,\infty)$, where the functional fit for the dimensionless pressure-volume curve is given as 
\begin{equation}
    \hat{p}(\overline{V}) = \sum_{m=0}^{N} a_m L_m(\overline{V}^{-1/3}),    \label{eq:Pbar_Vbar_Laguerre}
\end{equation}
and $a_m$ are the fitting coefficients.
Given that $L_m(0)=1$, the cavitation pressure is determined by the sum of the coefficients as
\begin{equation}
    \overline{p}_{\rm c}  = \lim_{\overline{V} \to \infty}\hat{p} = \sum_{m=0}^{N} a_m
    \label{eq:Pbar_Vbar_inf}
\end{equation}
The root-mean-square error (RMSE) of the fitted curve is defined as
\begin{equation}
\mathrm{RMSE} =
\sqrt{\frac{1}{n}\sum_{i=1}^{n}\left(\overline{p}_i - \hat{p}_i\right)^2}
\end{equation}
where $\hat{p}_i \coloneqq \hat{p}(\overline{V}_i)$ is the pressure predicted by the fitted model, $i$ denotes the index of the data points $(\overline{p}_i, \overline{V}_i)$ in the discrete numerical dataset, and $n$ is the number of data points in the dataset.\\

\begin{figure}[h]
    \centering
    \begin{subfigure}[t]{0.5\linewidth}
    \begin{overpic}[width=\linewidth]{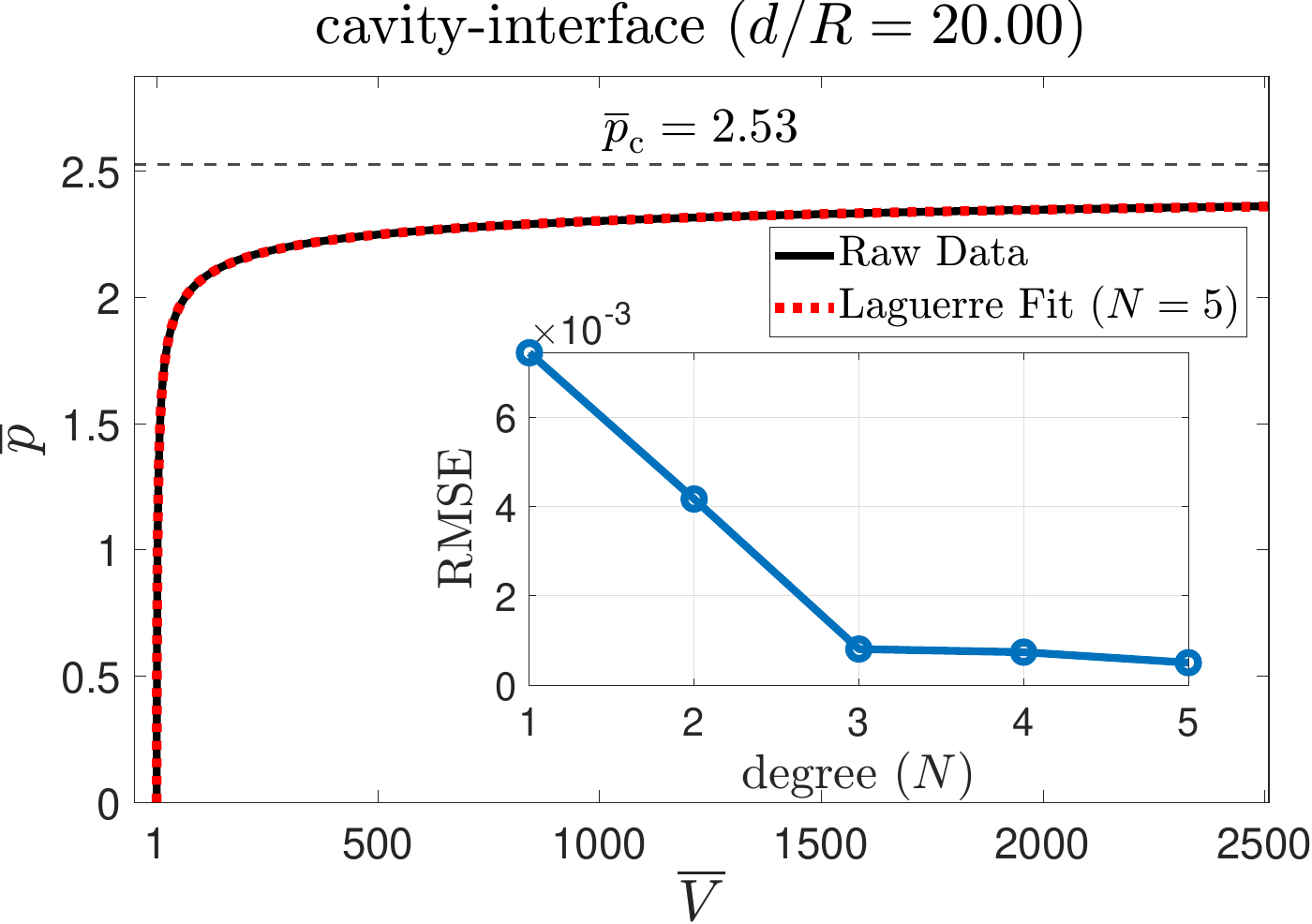}
    \put(0,66){\text{(a)}}
    \end{overpic}
    \phantomcaption
    \label{fig:fit_dR20_fixed}
    \end{subfigure}
    \hfill    
    \begin{subfigure}[t]{0.49\linewidth}
    \begin{overpic}[width=\linewidth]{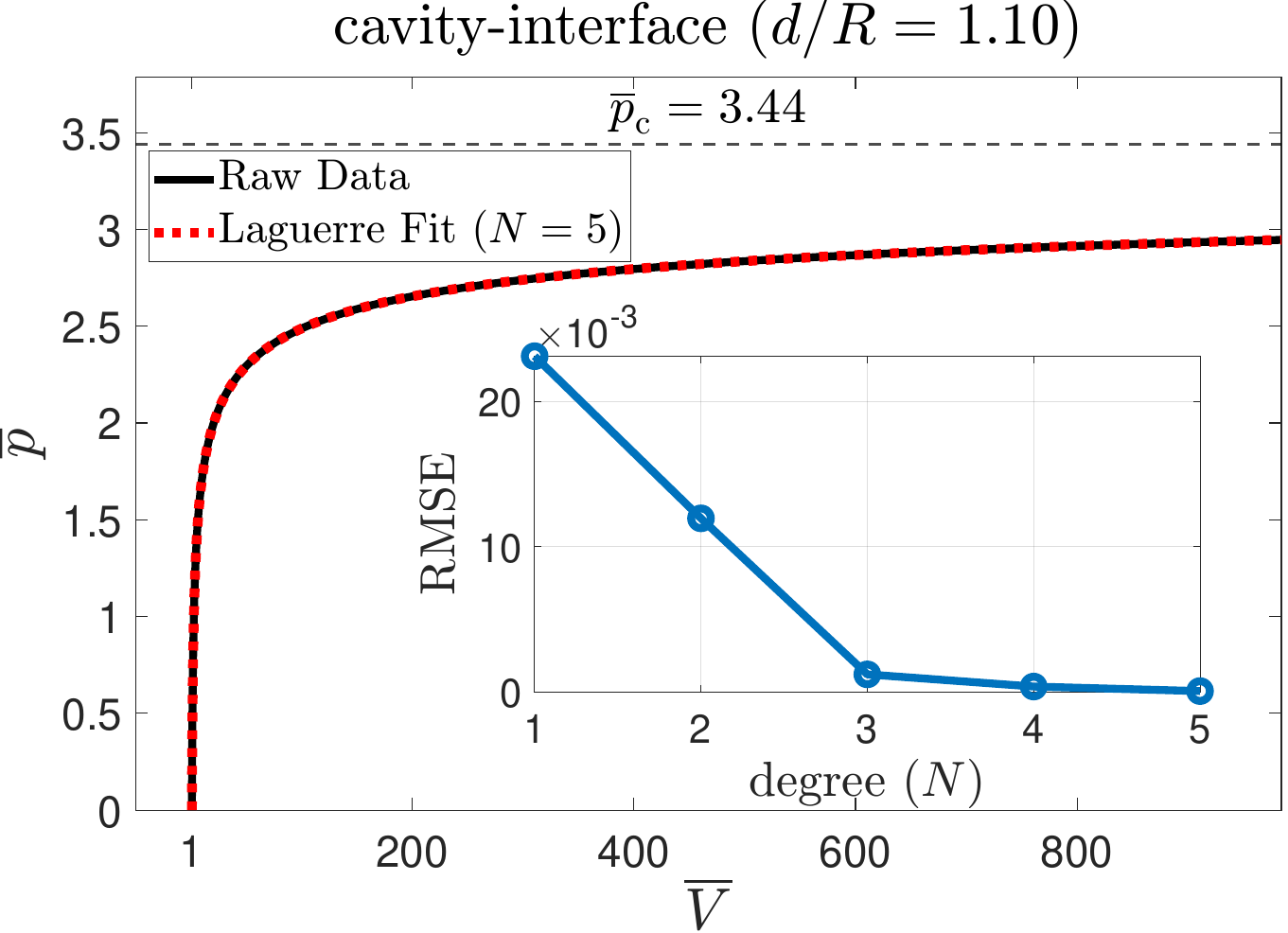}
    \put(0,68){\text{(b)}}
    \end{overpic}
    \phantomcaption
    \label{fig:fit_dR1_1_fixed}  
    \end{subfigure}
    \begin{subfigure}[t]{0.5\linewidth}
    \begin{overpic}[width=\linewidth]{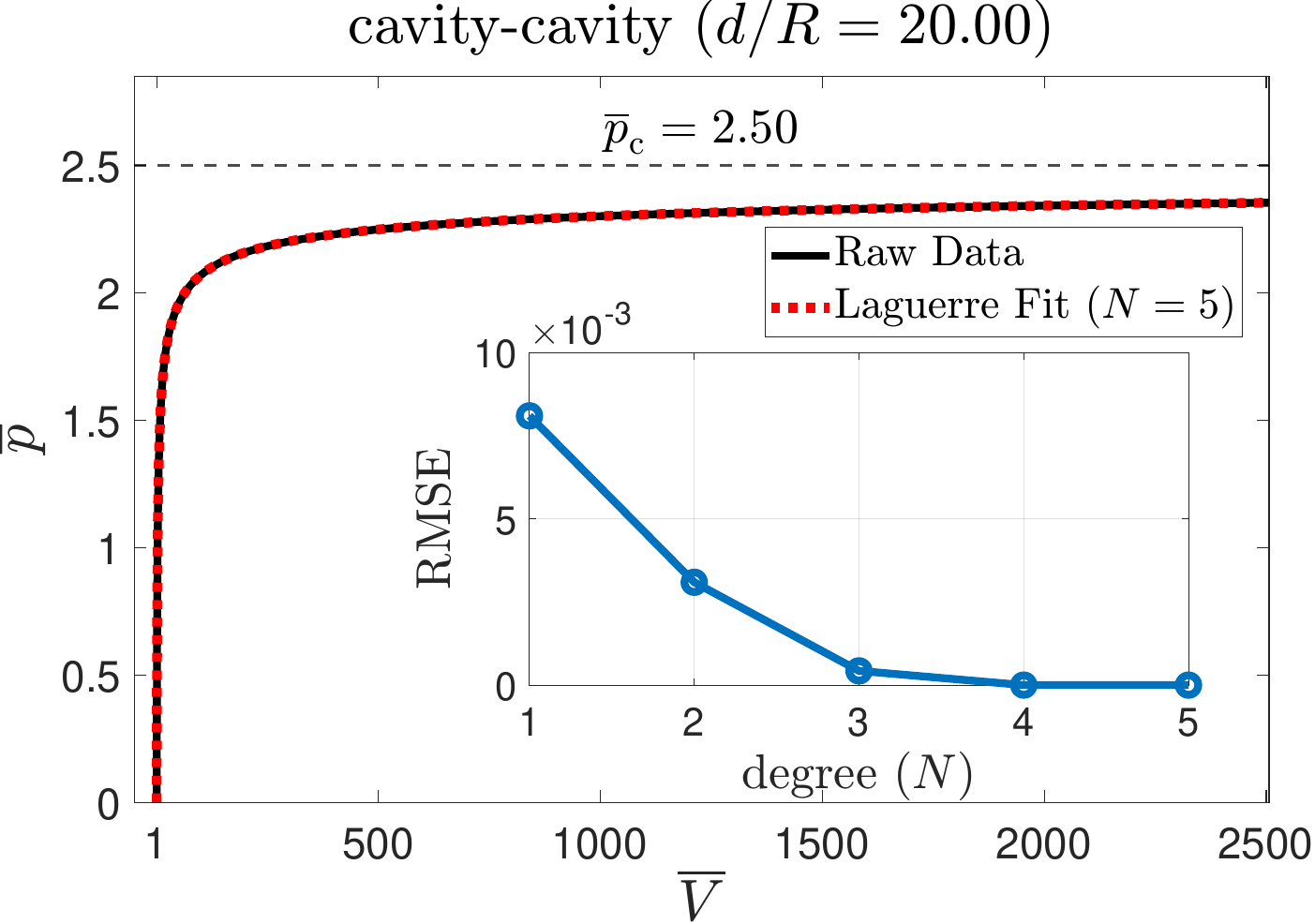}
    \put(0,66){\text{(c)}}
    \end{overpic}
    \phantomcaption
    \label{fig:fit_dR20}  
    \end{subfigure}
    \begin{subfigure}[t]{0.49\linewidth}
    \begin{overpic}[width=\linewidth]{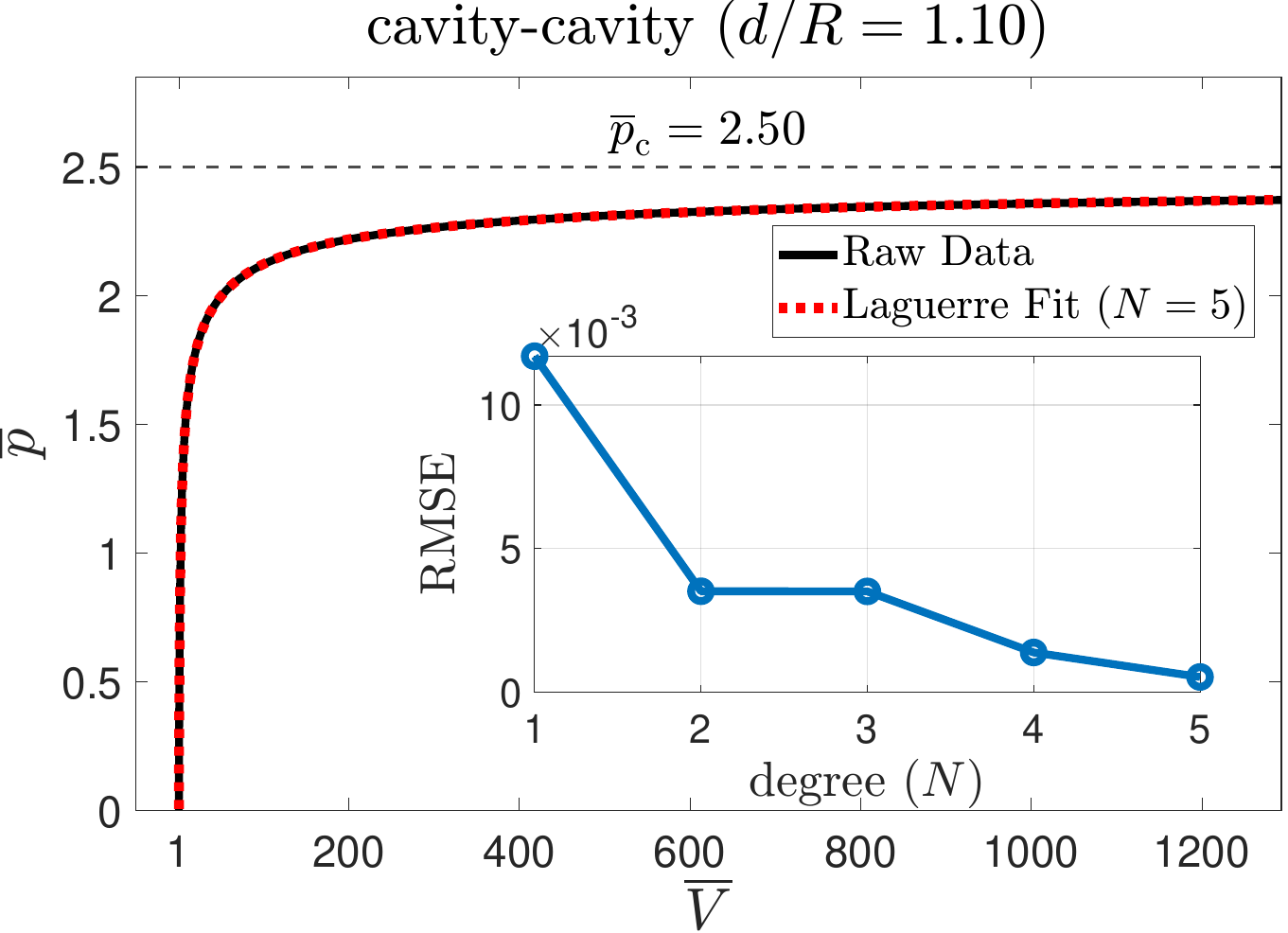}
    \put(0,68){\text{(d)}}
    \end{overpic}
    \phantomcaption
    \label{fig:fit_dR1_1}  
    \end{subfigure}
    \vspace{-15pt}
    \caption{Fitting procedure results for: (a) cavity-interface, $d/R = 20$ (b) cavity-interface, $d/R = 1.1$ (c) cavity-cavity, $d/R = 20$ and (d) cavity-cavity, $d/R = 1.1$.}
\end{figure}

We find that the RMSE decreases with increasing number of fitting terms $N+1$, however for $N>5$, the condition number of the matrix associated with the linear equation system to solve for the least squares fitting of the parameters $a_m$ becomes poor, reflecting over-parameterization and poor numerical stability of the fitted coefficients. Consequently, we cap the value of $N$ at 5, which gives satisfactory fit (low RMSE) for all $d/R$ values and report the average of the cavitation pressures obtained using the fits for $N=4$ and $N=5$ (the choice of $N\geq4$ covers all the terms in eq.~\eqref{eq:Pc_analytical}).
Further, we note that using a regular polynomial basis instead of the Laguerre polynomial basis yields identical results reflecting the robustness of the fitted cavitation pressure.
The fitting procedure is illustrated for two different $d/R$ values, for the cavity-interface interaction problem in Fig.~\ref{fig:fit_dR20_fixed}-\ref{fig:fit_dR1_1_fixed}, and for the cavity-cavity interaction problem in Fig.~\ref{fig:fit_dR20}-\ref{fig:fit_dR1_1}.

\section{Conclusion} \label{sec:conclusion}

We consider two extensions of the classical cavitation problem to analyse the effects of a rigid interface and, separately, another identical cavity on the critical pressure for cavitation in an incompressible, infinite neo-Hookean medium.
Through axisymmetric finite element simulations and a curve-fitting based (Laguerre polynomials basis functions) extrapolation procedure, we calculate the cavitation pressure as a function of the cavity-interface distance and inter-cavity half-distance, respectively.\\

For a cavity near a rigid interface, the cavitation pressure decreases monotonically with increasing cavity–interface spacing, with values ranging between {$p_{\rm c} \approx 3.44 \mu$} for small $d/R~(=1.1)$  and bulk cavitation value of $p_{\rm c, bulk} = 2.5\mu$ for large $d/R$. In cavity-cavity interaction case, the cavitation pressure is non-monotonic as a function of inter-cavity half-distance, with a peak cavitation pressure {$p_{\rm c} \approx 2.67\mu$} exhibited around $d/R\approx3.8$. The noticeable enhancement effect in both cases, with respect to the classical bulk cavitation pressure, indicates the importance of taking into account the presence of interface and nearby cavities, which is in turn likely to affect the mechanical response and failure in a variety of engineering and biological systems.

\appendix

\section{Mesh Sensitivity Analysis} \label{sec_appendix:mesh}
\label{app:convergence}

\begin{figure}[h]
    \centering
    \includegraphics[width=1\linewidth]{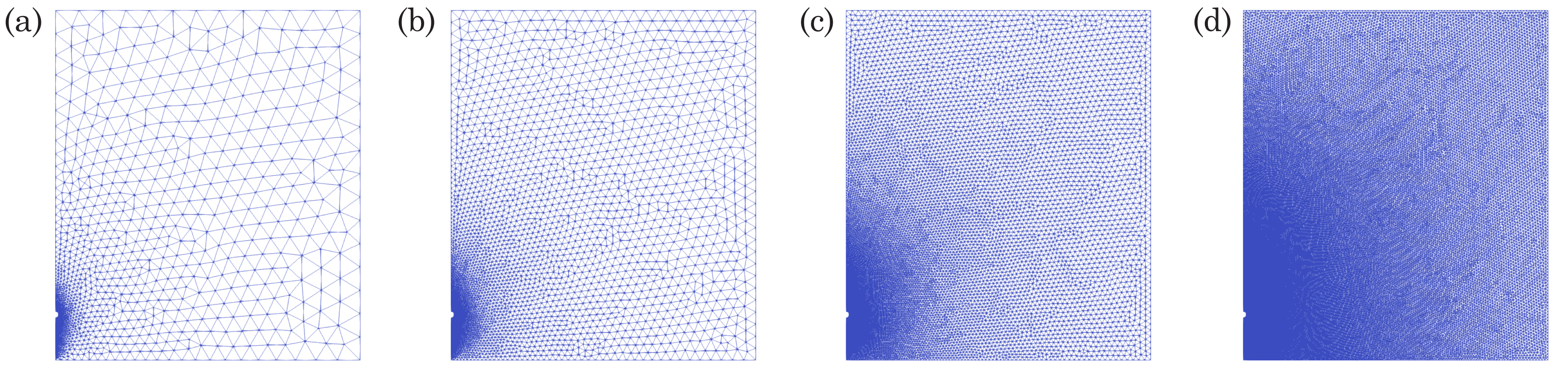}
    \caption{Different levels of adaptive refinement for $d/R=7.5$: (a) coarse (b) medium (c) fine (d) finest.}
    \label{fig:mesh}
\end{figure}

Four levels of mesh refinement were tested for a representative value of $d/R = 7.5$  as shown in Figure~\ref{fig:mesh}.
\noindent The remote pressure at the top boundary is calculated as 
\begin{equation}
    p=p_{\mathrm{remote},\mathrm{ave}}=\frac{\int_{X_2=L+d} \,(\boldsymbol{\sigma}\mathbf{n}. \mathbf{n})\,\mathrm{d}a(\mathbf{X})}{\int_{X_2=L+d}\,\mathrm{d}a(\mathbf{X})}
\end{equation}
where $\boldsymbol{\sigma}$ is the Cauchy stress tensor, $\mathbf{n}$ is the outward unit normal in the deformed configuration, $\mathbf{X}$ is the position vector of material points in the undeformed reference configuration, and $\mathrm{d}a$ is a deformed area of a differential surface element which is defined as
\begin{equation}
    \mathrm{d}a = 2 \pi r_0\,J \|\boldsymbol{\mathrm{F}}^{-T}\boldsymbol{\mathrm{N}}
    \|\,\mathrm{d}S
    \label{eq:da}
\end{equation}
where $r_0$ is the undeformed coordinate in $X_1$ direction, $\boldsymbol{\mathrm{F}}$ is the deformation gradient with $J=\mathrm{det}\boldsymbol{\mathrm{F}}(\approx 1)$, the outward unit normal in the reference configuration is $\boldsymbol{\mathrm{N}}$, and $\mathrm{d}S$ is the length of the differential arc element in the 2D axisymmetric representation. Moreover, the deformed volume of the cavity is determined as
\begin{equation}
    V = \int_{X_1^2+(X_2-d)^2=R^2} \pi r^2 \left(-n_1\right) \, \mathrm{d}s(\mathbf{X})
\end{equation}
where $r$ is the deformed coordinate in $X_1$ direction, $n_1$ is the component of the outward unit normal $\boldsymbol{\mathrm{n}}$ in $X_1$ direction, and $\mathrm{d}s$ is the deformed length of the differential arc element in the 2D axisymmetric representation. The deformed and undeformed arc lengths in the 2D axisymmetric plane are related through the relation
\begin{equation}
    \boldsymbol{\mathrm{n}}\, \mathrm{d}s = \textrm{det}(\boldsymbol{\mathrm{F}}_{\rm 2D}) \boldsymbol{\mathrm{F}}_{\rm 2D}^{-T}\boldsymbol{\mathrm{N}} \,\mathrm{d}S.
\end{equation}
where $\boldsymbol{\mathrm{F}}_{\rm 2D}$ is the deformation gradient in the 2D axisymmetric plane related to the full deformation gradient as follows,
\begin{equation}
\begin{aligned} \boldsymbol{\mathrm{F}} = \boldsymbol{\mathrm{F}}_{\rm 2D} + \frac{r}{r_0} \hat{e}_3 \otimes \hat{e}_3
\end{aligned}
\end{equation}
where $\hat{e}_3$ is a unit vector along $X_3$ axis.\\

\begin{figure}[h]
    \centering
    \begin{subfigure}[t]{0.493\linewidth}
    \begin{overpic}[width=\linewidth]{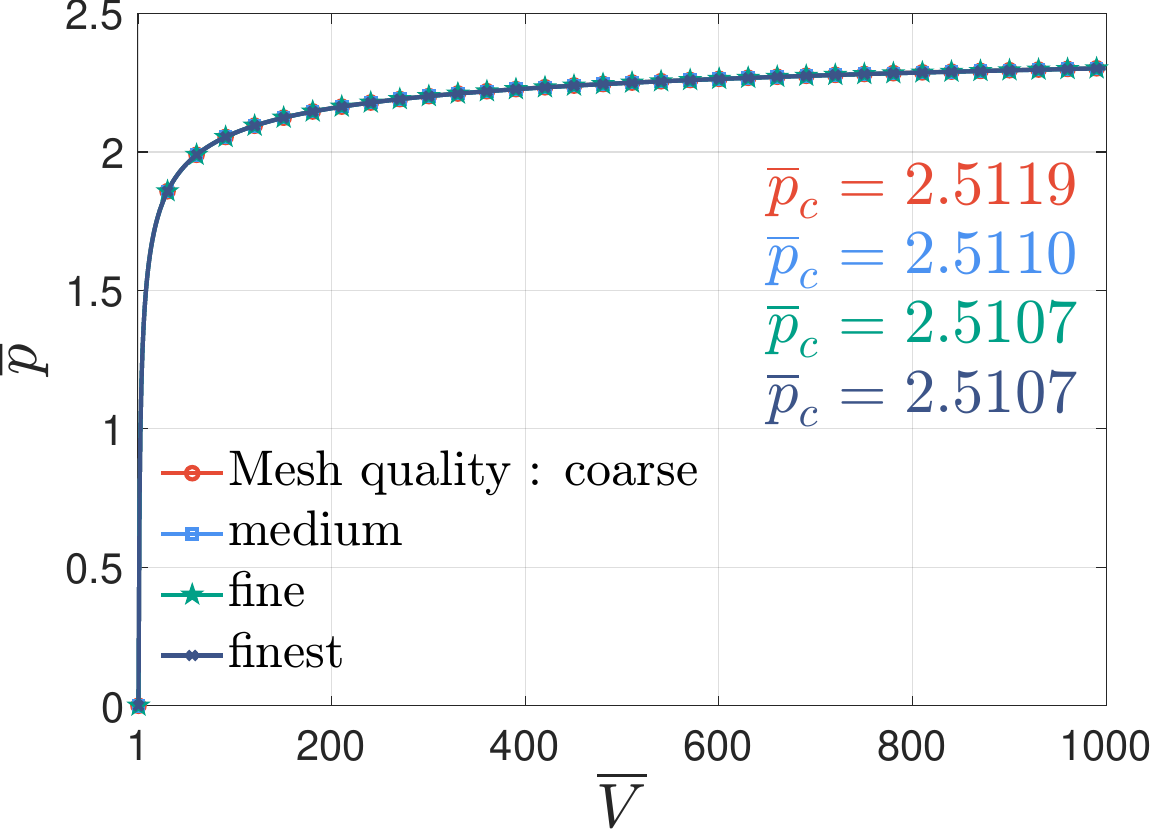}
    \put(0,73){\text{(a)}}
    \end{overpic}
    \phantomcaption
    \label{fig:remote_p}
    \end{subfigure}
    \hfill    
    \begin{subfigure}[t]{0.5\linewidth}
    \begin{overpic}[width=\linewidth]{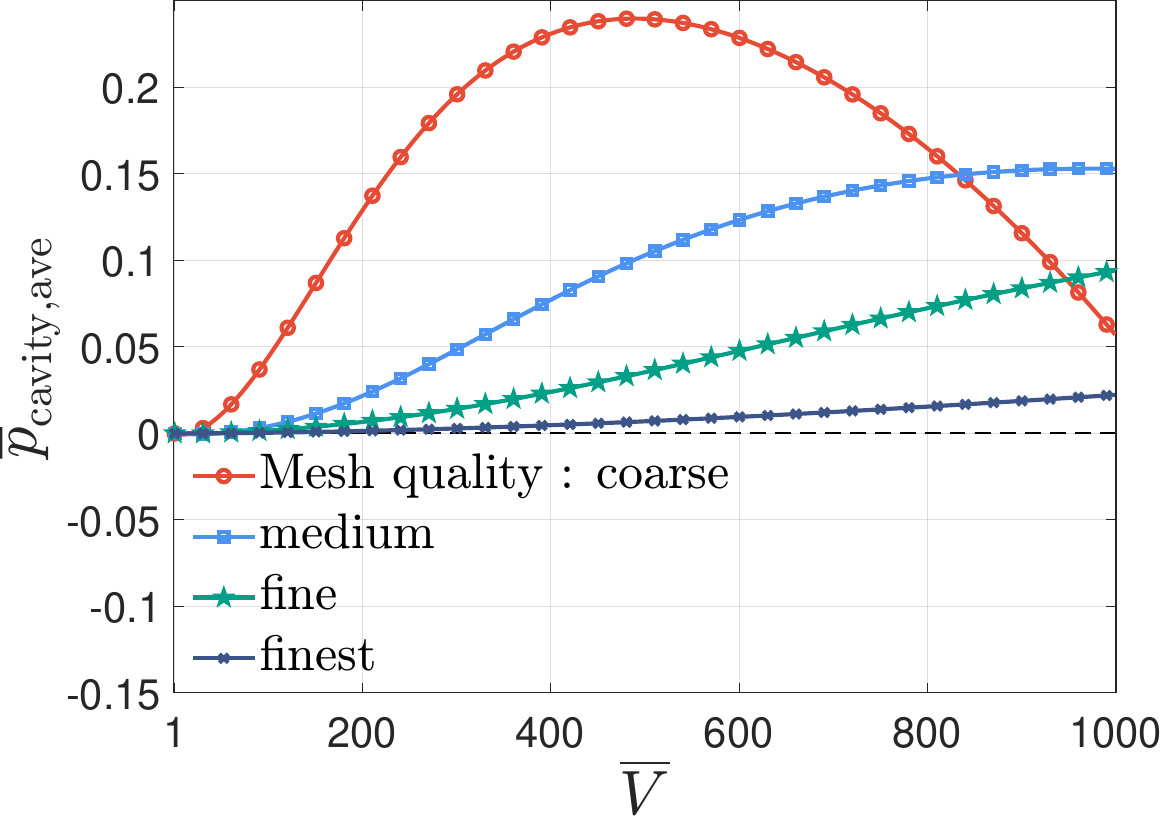}
    \put(0,72){\text{(b)}}
    \end{overpic}
    \phantomcaption
    \label{fig:cavity_p}  
    \end{subfigure}
    \vspace{-15pt}
    \caption{Cavity-cavity interaction problem for $d/R=7.5$ and different levels of mesh refinement:  (a) dimensionless pressure-volume curves, and (b) average dimensionless cavity pressure against dimensionless volume.}
\end{figure}

The dimensionless pressure vs volume curves obtained
for the different levels of mesh refinement are shown in Fig.~\ref{fig:remote_p} for the cavity-cavity interaction case. In addition, the asymptotic cavitation pressure is determined and shown in the figure using the corresponding colours. There is minimal change in the critical pressure estimate with mesh refinement and the remote pressure is well estimated even with a `coarse' mesh.

\noindent Next we check the traction free boundary condition at the cavity surface. The average pressure in the cavity is calculated as
\begin{equation}
    p_{\mathrm{cavity},\mathrm{ave}}=\frac{\int_{X_1^2+(X_2-d)^2=R^2} (\boldsymbol{\sigma}\mathbf{n}. \mathbf{n})\,\mathrm{d}a(\mathbf{X})}{\int_{X_1^2+(X_2-d)^2=R^2}\mathrm{d}a(\mathbf{X})}
\end{equation}
The average cavity pressure is presented in Fig.~\ref{fig:cavity_p} as a function of dimensionless deformed cavity volume, for the different levels of mesh refinement. The ``finest" mesh provides an acceptable accuracy in satisfying the natural boundary condition at the cavity, with an error of less than one percent. All results shown in the manuscript are using the finest level of mesh refinement.

 \bibliographystyle{elsarticle-num}
 \bibliography{cas-refs}

@article{kothari2020effect,
  title={Effect of elasticity on phase separation in heterogeneous systems},
  author={Kothari, Mrityunjay and Cohen, Tal},
  journal={Journal of the Mechanics and Physics of Solids},
  volume={145},
  pages={104153},
  year={2020},
  publisher={Elsevier}
}

@article{kothari2023crucial,
  title={The crucial role of elasticity in regulating liquid--liquid phase separation in cells},
  author={Kothari, Mrityunjay and Cohen, Tal},
  journal={Biomechanics and Modeling in Mechanobiology},
  volume={22},
  number={2},
  pages={645--654},
  year={2023},
  publisher={Springer}
}

@article{zhang2021morphogenesis,
  title={Morphogenesis and cell ordering in confined bacterial biofilms},
  author={Zhang, Qiuting and Li, Jian and Nijjer, Japinder and Lu, Haoran and Kothari, Mrityunjay and Alert, Ricard and Cohen, Tal and Yan, Jing},
  journal={Proceedings of the National Academy of Sciences},
  volume={118},
  number={31},
  pages={e2107107118},
  year={2021},
  publisher={National Academy of Sciences}
}

@article{nijjer2023biofilms,
  title={Biofilms as self-shaping growing nematics},
  author={Nijjer, Japinder and Li, Changhao and Kothari, Mrityunjay and Henzel, Thomas and Zhang, Qiuting and Tai, Jung-Shen B and Zhou, Shuang and Cohen, Tal and Zhang, Sulin and Yan, Jing},
  journal={Nature Physics},
  volume={19},
  number={12},
  pages={1936--1944},
  year={2023},
  publisher={Nature Publishing Group UK London}
}

@article{bonavia2026nonlinear,
  title={On the nonlinear Eshelby inclusion problem and its isomorphic growth limit},
  author={Bonavia, Joseph E and Chockalingam, S and Cohen, Tal},
  journal={Mathematics and Mechanics of Solids},
  volume={31},
  number={1},
  pages={140--171},
  year={2026},
  publisher={Sage Publications Sage UK: London, England}
}

@article{hang1992cavitation,
  title={Cavitation in elastic and elastic-plastic solids},
  author={Hou, Hang-Sheng and Abeyaratne, Rohan},
  journal={Journal of the Mechanics and Physics of Solids},
  volume={40},
  number={3},
  pages={571--592},
  year={1992},
  publisher={Elsevier}
}

@article{antman1987remarkable,
  title={The remarkable nature of radially symmetric equilibrium states of aeolotropic nonlinearly elastic bodies},
  author={Antman, Stuart S and Negr{\'o}n-Marrero, Pablo V},
  journal={Journal of Elasticity},
  volume={18},
  number={2},
  pages={131--164},
  year={1987},
  publisher={Springer}
}

@article{cohen2015dynamic,
  title={Dynamic cavitation and relaxation in incompressible nonlinear viscoelastic solids},
  author={Cohen, Tal and Molinari, Alain},
  journal={International Journal of Solids and Structures},
  volume={69},
  pages={544--552},
  year={2015},
  publisher={Elsevier}
}

@article{kumar2017some,
  title={Some remarks on the effects of inertia and viscous dissipation in the onset of cavitation in rubber},
  author={Kumar, Aditya and Aranda-Iglesias, Damian and Lopez-Pamies, Oscar},
  journal={Journal of Elasticity},
  volume={126},
  number={2},
  pages={201--213},
  year={2017},
  publisher={Springer}
}

@article{pericak1988nonuniqueness,
  title={Nonuniqueness for a hyperbolic system: cavitation in nonlinear elastodynamics},
  author={Pericak-Spector, KA and Spector, Scott J},
  journal={Archive for Rational Mechanics and Analysis},
  volume={101},
  number={4},
  pages={293--317},
  year={1988},
  publisher={Springer}
}

@article{meynard1992existence,
  title={Existence and nonexistence results on the radially symmetric cavitation problem},
  author={Meynard, Fran{\c{c}}ois},
  journal={Quarterly of applied mathematics},
  volume={50},
  number={2},
  pages={201--226},
  year={1992}
}

@article{horgan1992void,
  title={Void nucleation and growth for compressible non-linearly elastic materials: an example},
  author={Horgan, CO},
  journal={International journal of solids and structures},
  volume={29},
  number={3},
  pages={279--291},
  year={1992},
  publisher={Elsevier}
}

@article{dollhofer2004surface,
  title={Surface energy effects for cavity growth and nucleation in an incompressible neo-Hookean material----modeling and experiment},
  author={Dollhofer, J and Chiche, A and Muralidharan, V and Creton, C and Hui, CY},
  journal={International Journal of Solids and Structures},
  volume={41},
  number={22-23},
  pages={6111--6127},
  year={2004},
  publisher={Elsevier}
}

@article{polignone1993cavitation,
  title={Cavitation for incompressible anisotropic nonlinearly elastic spheres},
  author={Polignone, Debra A and Horgan, Cornelius O},
  journal={Journal of Elasticity},
  volume={33},
  number={1},
  pages={27--65},
  year={1993},
  publisher={Springer}
}

@article{horgan1995cavitation,
    author = {Horgan, C. O. and Polignone, D. A.},
    title = {Cavitation in Nonlinearly Elastic Solids: A Review},
    journal = {Applied Mechanics Reviews},
    volume = {48},
    number = {8},
    pages = {471-485},
    year = {1995},
    month = {08},
}

@article{STUART198533,
title = {Radially symmetric cavitation for hyperelastic materials},
journal = {Annales de l'Institut Henri Poincaré C, Analyse non linéaire},
volume = {2},
number = {1},
pages = {33-66},
year = {1985},
issn = {0294-1449},
author = {C.A. Stuart},
}

@article{lopez2011cavitationI,
  title={Cavitation in elastomeric solids: I—A defect-growth theory},
  author={Lopez-Pamies, Oscar and Idiart, Martin I and Nakamura, Toshio},
  journal={Journal of the Mechanics and Physics of Solids},
  volume={59},
  number={8},
  pages={1464--1487},
  year={2011},
  publisher={Elsevier}
}

@book{Anand2026,
  author    = {Anand, Lallit and Stewart, Eric M. and Chester, Shawn A.},
  title     = {Introduction to Coupled Theories in Solid Mechanics},
  publisher = {Oxford University Press},
  address   = {Oxford},
  year      = {2026},
  isbn      = {9780198986218},
  series    = {Oxford Graduate Texts}
}

@article{huang1991cavitation,
  title={Cavitation instabilities in elastic-plastic solids},
  author={Huang, Y and Hutchinson, JW and Tvergaard, V},
  journal={Journal of the Mechanics and Physics of Solids},
  volume={39},
  number={2},
  pages={223--241},
  year={1991},
  publisher={Elsevier}
}

@article{barney2020cavitation,
  title={Cavitation in soft matter},
  author={Barney, Christopher W and Dougan, Carey E and McLeod, Kelly R and Kazemi-Moridani, Amir and Zheng, Yue and Ye, Ziyu and Tiwari, Sacchita and Sacligil, Ipek and Riggleman, Robert A and Cai, Shengqiang and others},
  journal={Proceedings of the National Academy of Sciences},
  volume={117},
  number={17},
  pages={9157--9165},
  year={2020},
  publisher={National Academy of Sciences}
}

@article{lopez2011cavitationII,
  title={Cavitation in elastomeric solids: II—Onset-of-cavitation surfaces for Neo-Hookean materials},
  author={Lopez-Pamies, Oscar and Nakamura, Toshio and Idiart, Mart{\'\i}n I},
  journal={Journal of the Mechanics and Physics of Solids},
  volume={59},
  number={8},
  pages={1488--1505},
  year={2011},
  publisher={Elsevier}
}

@article{sivaloganathan1986uniqueness,
  title={Uniqueness of regular and singular equilibria for spherically symmetric problems of nonlinear elasticity},
  author={Sivaloganathan, Jeyabal},
  journal={Archive for Rational Mechanics and Analysis},
  volume={96},
  number={2},
  pages={97--136},
  year={1986},
  publisher={Springer-Verlag Berlin/Heidelberg}
}

@article{horgan1986bifurcation,
  title={A bifurcation problem for a compressible nonlinearly elastic medium: growth of a micro-void},
  author={Horgan, CO and Abeyaratne, R},
  journal={Journal of Elasticity},
  volume={16},
  number={2},
  pages={189--200},
  year={1986},
  publisher={Springer}
}

@article{li2025cylindrical,
  title={Cylindrical cavity expansion for characterizing mechanical properties of soft materials},
  author={Li, Jian and Xie, Zihao and Varner, Hannah and Chockalingam, S and Cohen, Tal},
  journal={Extreme Mechanics Letters},
  volume={77},
  pages={102343},
  year={2025},
  publisher={Elsevier}
}

@article{chockalingam2021probing,
  title={Probing local nonlinear viscoelastic properties in soft materials},
  author={Chockalingam, S and Roth, Christine and Henzel, Thomas and Cohen, Tal},
  journal={Journal of the Mechanics and Physics of Solids},
  volume={146},
  pages={104172},
  year={2021},
  publisher={Elsevier}
}

@article{chockalingam2024large,
  title={A large deformation theory for coupled swelling and growth with application to growing tumors and bacterial biofilms},
  author={Chockalingam, S and Cohen, Tal},
  journal={Journal of the Mechanics and Physics of Solids},
  volume={187},
  pages={105627},
  year={2024},
  publisher={Elsevier}
}

@article{li2022nonlinear,
  title={Nonlinear inclusion theory with application to the growth and morphogenesis of a confined body},
  author={Li, Jian and Kothari, Mrityunjay and Chockalingam, S and Henzel, Thomas and Zhang, Qiuting and Li, Xuanhe and Yan, Jing and Cohen, Tal},
  journal={Journal of the Mechanics and Physics of Solids},
  volume={159},
  pages={104709},
  year={2022},
  publisher={Elsevier}
}

@article{chockalingam2022thermo,
  title={Thermo-chemo-mechanically coupled cavity dynamics and the emergence of multi-phase bursts},
  author={Chockalingam, S and Lem, J and Cohen, T},
  journal={Proceedings of the Royal Society A: Mathematical, Physical and Engineering Sciences},
  volume={478},
  number={2264},
  year={2022},
  publisher={The Royal Society}
}

@article{henzel2022reciprocal,
  title={A reciprocal theorem for finite deformations in incompressible bodies},
  author={Henzel, Thomas and Senthilnathan, Chockalingam and Cohen, Tal},
  journal={arXiv preprint arXiv:2201.08338},
  year={2022}
}

@phdthesis{senthilnathan2024understanding,
  title={Understanding the mechanics of growth: A large deformation theory for coupled swelling-growth and morphogenesis of soft biological systems},
  author={Senthilnathan, Chockalingam},
  year={2024},
  school={Massachusetts Institute of Technology}
}

@article{Saeedi2025,
  title={Push and Pull: Elastic Interaction Between Pressurized Spherical Cavities in Nonlinear Elastic Media},
  author={Saeedi, Ali and Kothari, Mrityunjay},
  journal={Journal of Applied Mechanics},
  volume={92},
  number={12},
  pages={121010},
  year={2025},
  publisher={American Society of Mechanical Engineers}
}

@article{Henzel2022,
    author = {Henzel, Thomas and Nijjer, Japinder and Chockalingam, S and Wahdat, Hares and Crosby, Alfred J and Yan, Jing and Cohen, Tal},
    title = {Interfacial cavitation},
    journal = {PNAS Nexus},
    volume = {1},
    year = {2022},
    pages = {pgac217},
}

@article{GentLindley1959,
  author    = {A. N. Gent and P. B. Lindley},
  title     = {Internal rupture of bonded rubber cylinders in tension},
  journal   = {Proceedings of the Royal Society of London. Series A, Mathematical and Physical Sciences},
  volume    = {249},
  number    = {1257},
  pages     = {195--205},
  year      = {1959},
}

@article{Ball1982,
  author    = {J. M. Ball},
  title     = {Discontinuous equilibrium solutions and cavitation in nonlinear elasticity},
  journal   = {Philosophical Transactions of the Royal Society of London. Series A, Mathematical and Physical Sciences},
  volume    = {306},
  number    = {1496},
  pages     = {557--611},
  year      = {1982},
}

@article{GentPark1984,
  author    = {A. N. Gent and B. Park},
  title     = {Failure processes in elastomers at or near a rigid spherical inclusion},
  journal   = {Journal of Materials Science},
  volume    = {19},
  number    = {6},
  pages     = {1947--1956},
  year      = {1984},
}

@article{cho1988cavitation,
  title={Cavitation in model elastomeric composites},
  author={Cho, K and Gent, AN},
  journal={Journal of materials science},
  volume={23},
  number={1},
  pages={141--144},
  year={1988},
  publisher={Springer}
}

@article{LopezPamies2009,
  author    = {Oscar Lopez-Pamies},
  title     = {Onset of cavitation in compressible, isotropic, hyperelastic solids},
  journal   = {Journal of elasticity},
  volume    = {94},
  number    = {2},
  pages     = {115--145},
  year      = {2009},
}

@article{LopezPamies2007,
  title={Homogenization-based constitutive models for porous elastomers and implications for macroscopic instabilities: {II}—Results},
  author={Lopez-Pamies, Oscar and Casta{\~n}eda, P Ponte},
  journal={Journal of the Mechanics and Physics of Solids},
  volume={55},
  number={8},
  pages={1702--1728},
  year={2007},
  publisher={Elsevier}
}

@article{Kang2018,
  author    = {Jingtian Kang and Changguo Wang and Huifeng Tan},
  title     = {Cavitation in inhomogeneous soft solids},
  journal   = {Soft Matter},
  volume    = {14},
  number    = {39},
  pages     = {7979--7986},
  year      = {2018},
}

@article{Li2025,
  author    = {Xuanhe Li and Brendan Unikewicz and S. Chockalingam and Hudson Borja da Rocha and Tal Cohen},
  title     = {Interfacial cavitation with surface tension: New insights into failure of particle reinforced polymers},
  journal   = {Journal of the Mechanics and Physics of Solids},
  volume    = {206},
  pages     = {106379},
  year      = {2026},
}

@article{Saeedi2D,
  title={Elastic interaction of pressurized cavities in hyperelastic media: Attraction and repulsion},
  author={Saeedi, Ali and Kothari, Mrityunjay},
  journal={Journal of Applied Mechanics},
  volume={92},
  number={5},
  pages={051008},
  year={2025},
  publisher={American Society of Mechanical Engineers}
}

@article{alnaes2015fenics,
  title={The FEniCS project version 1.5},
  author={Aln{\ae}s, Martin and Blechta, Jan and Hake, Johan and Johansson, August and Kehlet, Benjamin and Logg, Anders and Richardson, Chris and Ring, Johannes and Rognes, Marie E and Wells, Garth N},
  journal={Archive of numerical software},
  volume={3},
  number={100},
  year={2015}
}

@book{logg2012automated,
  title={Automated solution of differential equations by the finite element method: The FEniCS book},
  author={Logg, Anders and Mardal, Kent-Andre and Wells, Garth},
  volume={84},
  year={2012},
  publisher={Springer Science \& Business Media}
}

@incollection{logg2012dolfin,
  title={DOLFIN: a C++/Python finite element library},
  author={Logg, Anders and Wells, Garth N and Hake, Johan},
  booktitle={Automated Solution of Differential Equations by the Finite Element Method: The FEniCS Book},
  pages={173--225},
  year={2012},
  publisher={Springer}
}

@article{geuzaine2009gmsh,
  title={Gmsh: A 3-D finite element mesh generator with built-in pre-and post-processing facilities},
  author={Geuzaine, Christophe and Remacle, Jean-Fran{\c{c}}ois},
  journal={International journal for numerical methods in engineering},
  volume={79},
  number={11},
  pages={1309--1331},
  year={2009},
  publisher={Wiley Online Library}
}
 
\end{document}